\documentclass{PoS}

\usepackage{lineno}

\usepackage[gen]{eurosym}  
\usepackage{wasysym}

\graphicspath{{figs/}}

\title{Search for new supernova remnant shells in the Galactic plane with H.E.S.S.}

\ShortTitle{Search for new SNR shells with H.E.S.S.}

\author{\speaker{G.~P{\"u}hlhofer}$^{,a}$\\
        \llap{$^a$} Institut f{\"u}r Astronomie und Astrophysik, Abteilung Hochenergieastrophysik, Kepler Center for Astro and Particle Physics, Eberhard Karls Universit{\"a}t, Sand 1, D 72076 T{\"u}bingen, Germany\\
        E-mail: \email{Gerd.Puehlhofer@astro.uni-tuebingen.de}}

\author{F.~Brun$^b$,  M.~Capasso$^a$, R.C.G.~Chaves$^c$, C.~Deil$^d$, A.~Djannati-Atai$^e$, A.~Donath$^d$, P.~Eger$^d$, D.~Gottschall$^a$, H.~Laffon$^f$, V.~Marandon$^d$, L.~Oakes$^g$, M.~Renaud$^c$, M.~Sasaki$^a$, R.~Terrier$^e$, J.~Vink$^h$,\\
\llap{$^b$} DSM/Irfu, CEA Saclay, F-91191 Gif-Sur-Yvette Cedex, France \\
\llap{$^c$} Laboratoire Univers et Particules de Montpellier, Universit\'e Montpellier 2, CNRS/IN2P3,  CC 72, Place Eug\`ene Bataillon, F-34095 Montpellier Cedex 5, France \\
\llap{$^d$} Max-Planck-Institut f{\"u}r Kernphysik, P.O. Box 103980, D 69029 Heidelberg, Germany\\
\llap{$^e$} APC, AstroParticule et Cosmologie, Universit\'{e} Paris Diderot, CNRS/IN2P3, CEA/Irfu, Observatoire de Paris, Sorbonne Paris Cit\'{e}, 10, rue Alice Domon et L\'{e}onie Duquet, 75205 Paris Cedex 13, France \\
\llap{$^f$}  Universit\'e Bordeaux, CNRS/IN2P3, Centre d'\'Etudes Nucl\'eaires de Bordeaux Gradignan, 33175 Gradignan, France \\
\llap{$^g$} Institut f\"ur Physik, Humboldt-Universit\"at zu Berlin, Newtonstr. 15, D 12489 Berlin, Germany \\
\llap{$^h$} GRAPPA, Anton Pannekoek Institute for Astronomy, University of Amsterdam,  Science Park 904, 1098 XH Amsterdam, The Netherlands 
}
\author{for the H.E.S.S. collaboration,}

\author{A.~Bamba$^i$\\
        \llap{$^i$} Department of Physics and Mathematics, Aoyama Gakuin University 5-10-1 Fuchinobe Chuo-ku, Sagamihara, Kanagawa 252-5258, Japan}

\abstract{Amongst the population of TeV $\gamma$-ray sources detected with the High Energy Stereoscopic System (H.E.S.S.) in the Galactic plane, clearly identified supernova remnant (SNR) shells constitute a small but precious source class. TeV-selected SNRs are prime candidates for sources of efficient cosmic-ray acceleration. In this work, we present new SNR candidates that have been identified in the entire H.E.S.S. phase I data set of the Galactic plane recorded over the past ten years. Identification with a known SNR shell candidate was successful for one new source, HESS\,J1534$-$571. In other cases, TeV-only shell candidates are challenging to firmly identify as SNRs due to their lack of detected non-thermal emission in lower energy bands. We will discuss how these objects may present an important link between young and evolved SNRs, since their shell emission may be dominated by hadronic processes.}

\FullConference{The 34th International Cosmic Ray Conference,\\
		30 July--6 August, 2015\\
		The Hague, The Netherlands}

\begin{document}

\section{Introduction}

During the past $\sim$10 years of operation, a significant fraction of the H.E.S.S.\footnote{H.E.S.S. (High Energy Stereoscopic System) is a system of Cherenkov telescopes operated in the Khomas Highlands in Namibia. H.E.S.S. phase I started in 2003 with four 12-m telescopes. H.E.S.S. phase II is defined through the additional fifth, 28-m telescope CT\,5 that has been operated together with the other telescopes starting with commissioning in 2012.} telescopes' observing time has been used to observe the Galactic plane, both through a dedicated survey and through pointed observations of selected targets or regions. The H.E.S.S. I telescopes, with a point spread function for reconstructed $\gamma$-rays of $\sim 0.05^{\circ}$ to $0.1^{\circ}$ (depending on analysis configuration) and a large field-of-view with $\diameter \sim 3^{\circ}$ flat $\gamma$-ray acceptance\footnote{The field-of-view of the H.E.S.S. I cameras for air showers corresponds to a circle with $\diameter = 5^{\circ}$.}, are ideally suited for the discovery and study of the typically $\lesssim 0.5^{\circ}$ extended TeV $\gamma$-ray sources seen in our Galaxy. A significant fraction of the known Galactic TeV source population has actually been detected with H.E.S.S.\footnote{\href{http://tevcat.uchicago.edu}{http://tevcat.uchicago.edu}} For a recent TeV $\gamma$-ray astronomy review see e.g. \cite{bib:rieger2013}.  

The largest fraction of the currently known Galactic TeV $\gamma$-ray sources consists of still unidentified objects. Amongst those, many are pulsar wind nebula (PWN) candidates. Sources are classified as PWN candidates if they are found in positional coincidence with pulsars that are energetic enough to have powered a nebula of outwards diffusing relativistic electrons. The TeV emission is interpreted to stem from these electrons, via Inverse Compton (IC) scattering off background photons. Ideally, an X-ray PWN is also detected at the pulsar position. The shape of the TeV emission in general does not permit a firm identification as a PWN, since a head-tail morphology such as in X-rays is not observed. Neither does the TeV spectrum alone permit an identification. Nevertheless, the largest fraction of identified Galactic sources are classified as PWNe.

It is, however, very likely that some of the unidentified TeV $\gamma$-ray sources are due to emission from particles accelerated in hitherto unknown SNRs. The {\em established} TeV SNRs are distinguished by their shell-like appearance and their TeV morphology matching their shell-like counterparts in radio and non-thermal X-rays: RXJ\,1713.7$-$3946, Vela Jr., RCW\,86, the remnant of SN\,1006\footnote{The TeV emission from SN\,1006 is in fact not shell-like but bipolar, matching however clearly the appearance in non-thermal X-rays.}, and HESS\,J1731$-$347. The TeV emission from Cassiopeia A is very likely also shell-dominated, but the TeV emission is unresolved. A couple of other TeV sources are likely associated with SNR processes as well, through the possible association of the TeV emission with molecular clouds close to or partially coincident with the SNRs. The underlying assumption is that emission from SNR-accelerated hadronic particles is boosted in these clouds. Examples comprise the TeV emission from W51C, IC\,443, W49B, CTB\,37A, and from HESS\,J1800$-$240 near W\,28, the latter likely representing the best example of particles that have entered non-shock-compressed clouds outside the SNR. Often, the firm identification of the TeV emission with this scenario is challenging, due to the statistics-limited TeV images that are to be correlated with sub-mm line data (tracing molecular gas) which provide moderate distance resolution and thus a large phase-space. 

The work presented here deals with the question of whether or not previously unidentified or new TeV $\gamma$-ray sources in the H.E.S.S.\ phase I Galactic plane data can be identified as SNRs. A spectral identification is not possible, and associations with molecular clouds are difficult due to the aforementioned problems. A possible approach is to look for shell-like morphological appearance of extended ($\gg 0.1^{\circ}$) TeV sources, compared to the typical Gaussian morphology of other known TeV sources of similar size. One established TeV source which falls into this category is HESS\,J1731$-$347, which was first classified as an unidentified TeV source \cite{bib:darkhess2008} and later confirmed as having a shell-type morphology using a much deeper TeV data set \cite{bib:1731hess2011}. Before the TeV shell identification, there was, however, already strong evidence that the TeV source is an SNR, from the detection of a radio SNR with position and angular size matching the TeV source \cite{bib:tian17132008}.

\section{Shell search methodology}

To search for shell-like sources (with expected resolvable angular scales on the order of $0.2^{\circ}$ to $1^{\circ}$) in statistics-limited sky maps, with a population of other TeV sources (with typical angular separation scales of degrees) as background, is challenging. Amongst the problems are the following two items.

\noindent {\em Choice of the regions of interest:}
It is often ambiguous if an emission region consists of one or several (at least in projection closeby) astrophysical objects. There is no unique decision criterion, and some choice on the assumed typical source morphology has to be made. Emission regions with extended shell morphologies may not necessarily be identified as a single source. As a consequence, an unbiased search can only be performed on a sky spatial grid, where at each fixed position a shell morphology is compared to a morphological null hypothesis.

\noindent {\em Choice of the null hypothesis:}
The null hypothesis of the morphology test needs to be chosen. In principle, one could perform Monte-Carlo simulations at each test position based on a training sample of typical TeV source morphologies. With this, one could evaluate what the likelihood is that by chance a circular morphology appears. Such analysis is however well beyond the scope of the presented work. 

In order to simplify, a centrally peaked source (represented here by a 2-dimensional Gaussian) is adopted as null hypothesis. The target morphology is an azimuthally symmetric, homoge\-neously-emitting 3-dimensional shell between minimum and maximum radii, respectively, projected onto sky coordinates. This morphology is a good zero-order approximation for the known TeV-emitting SNR shells (marginal for SN\,1006). Concerning the choice of the regions of interest, a two-step approach was adopted: First, the search was conducted on a predefined {\em grid of sky coordinates}, using the H.E.S.S. Galactic plane survey (HGPS) sky maps \cite{bib:HGPSICRC2015}, with a binning of $0.02^{\circ}\times 0.02^{\circ}$ equidistant spacing on the sky. At each grid position, the best-fit shell was compared to the best-fit null-hypothesis morphology (Gaussian). Second, the best candidates from the grid search were then evaluated {\em on individual source basis}, i.e.\ on sky maps on which only one selected emission region of interest is present at a time. Here, the best-fit positions for the shell and null-hypothesis morphology (Gaussian) were left free and thus possibly different from each other.

Given the issues of likely source confusion at many locations and the necessary choice of tested morphologies, it is evident that the search cannot be complete to any sensitivity level. The main goal was to identify obvious shell candidates, but no statistical assessment on the likelihood that the source is indeed a SNR, or on the likelihood of how many candidate shells are not found by the method, was performed. Thus, 

\begin{itemize}
  \item a H.E.S.S. source with reasonably significant shell morphology, identified by this search, is considered a TeV SNR candidate; there are other astrophysical objects that potentially may appear shell-like and are potential TeV $\gamma$-ray emitters, such as superbubbles or wind-blown cavities into which hadronic particles are diffusing, but given the experience from the currently identified SNRs and the lack of identified alternative scenarios this approach seems feasible; 
  \item a H.E.S.S. shell (or shell candidate if the significance for a shell from TeV data alone is marginal) is only identified as an SNR if an SNR counterpart or strong SNR counterpart candidate in another wavelength can be associated with the TeV source, through matching morphology, given the lack of possibility of an independent confirmation from the TeV data alone (such as variability and spectrum).
\end{itemize}

In the following, two prominent examples that emerged from the search for new SNR shells are presented. Only data from H.E.S.S. phase I were used; some data stem from the commissioning phase of CT\,5, but CT\,5 data were not yet included in regular data taking.\footnote{The hardware trigger configuration of phase II is such that software-wise exclusion of CT\,5 from analysis yields data with equivalent sensitivity as H.E.S.S. phase I data.} The H.E.S.S. analysis is still preliminary. Refined results will be presented at the conference.

\begin{figure}[t]
    \centering
	\includegraphics[width=0.495\textwidth]{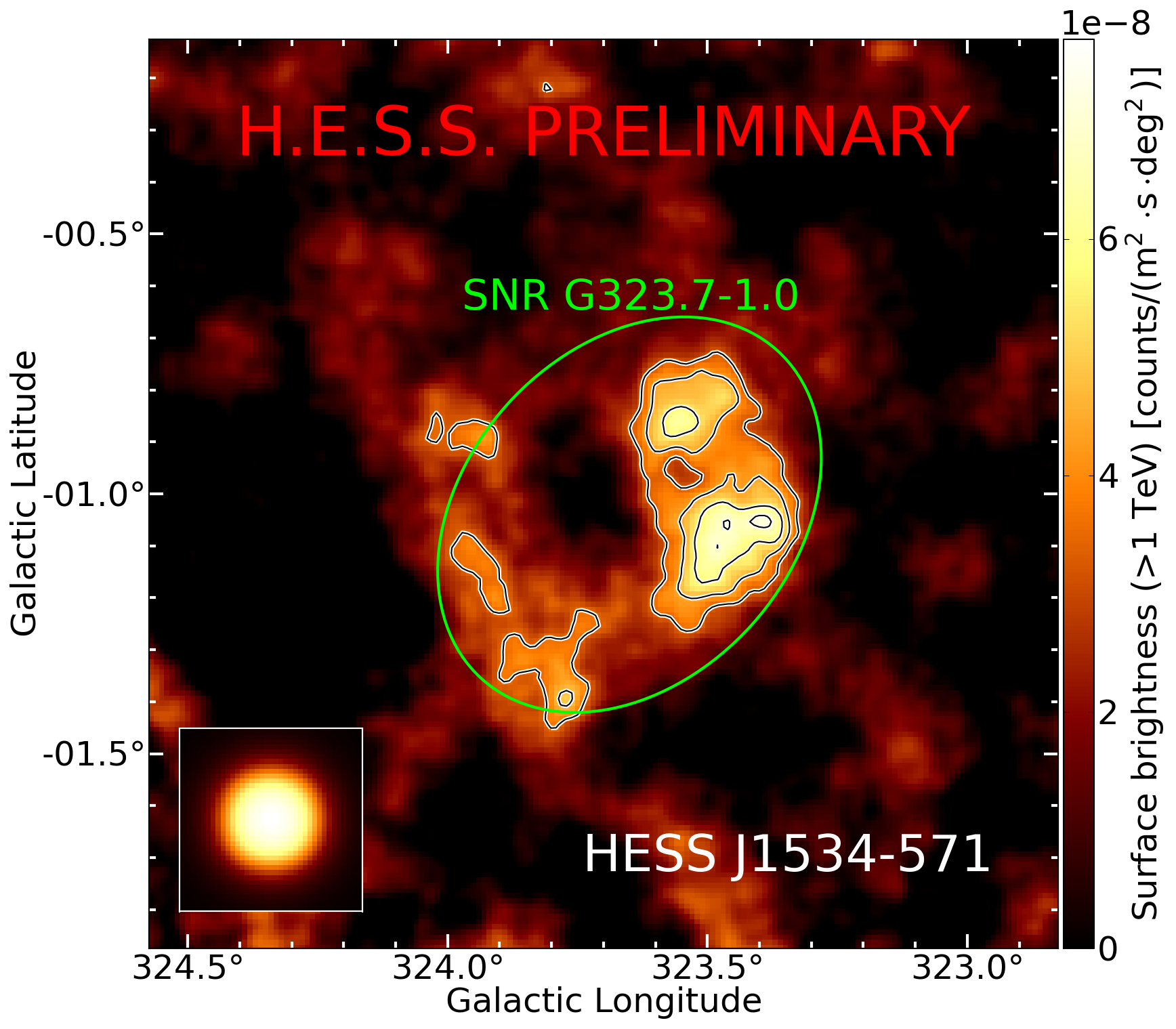}
	\includegraphics[width=0.495\textwidth]{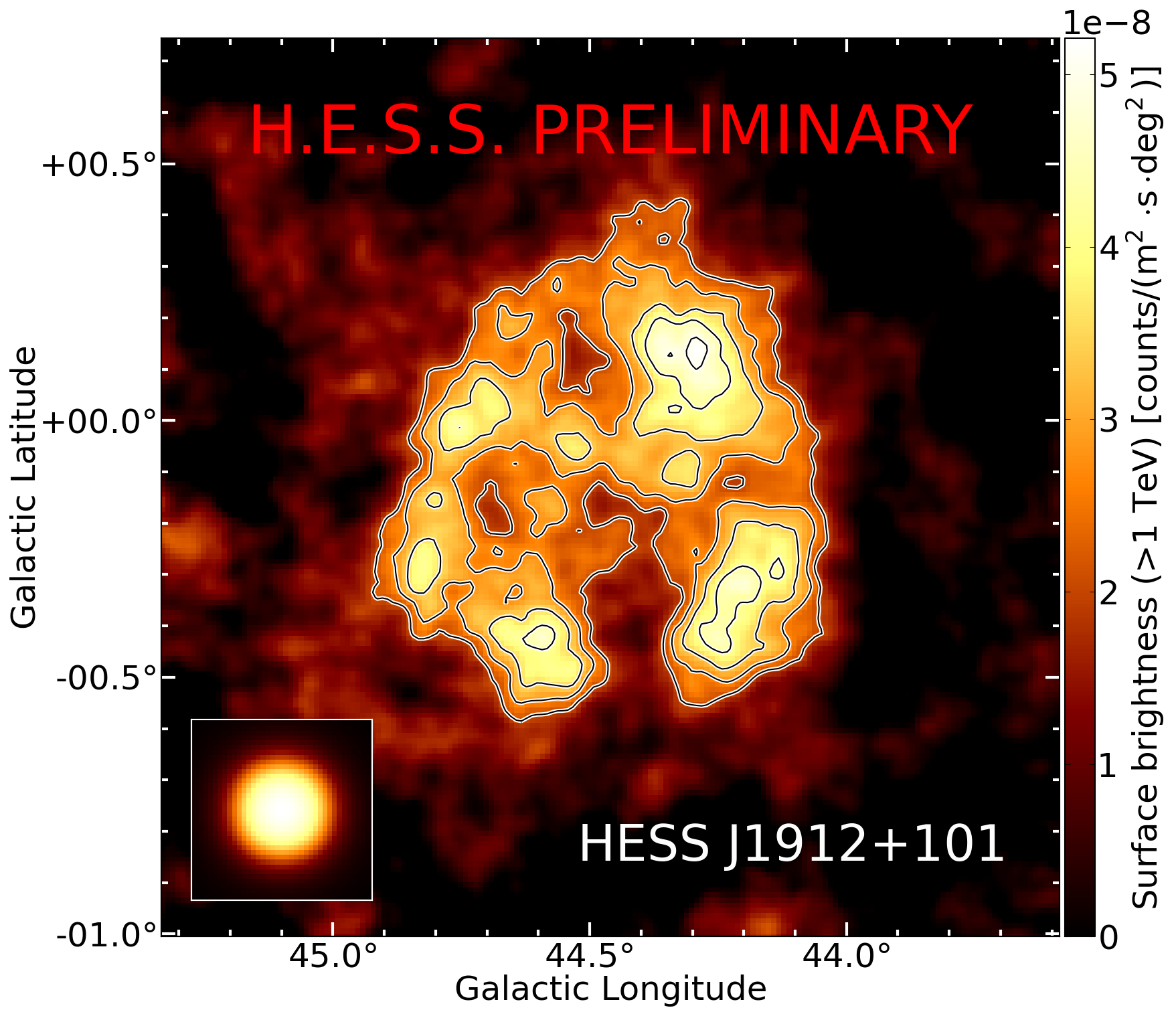}
	\caption{Left: TeV surface brightness map of HESS\,J1534$-$571, derived from all excess counts in circles of (oversampling) radius $0.1^{\circ}$ and converted into a count rate above 1\,TeV, assuming a source spectral photon index of $\Gamma = 2.3$, as indicated by a preliminary spectral analysis. The map is then for better representation additionally mildly smoothed with a 2-dimensional Gaussian filter with $\sigma = 0.01^{\circ}$. Overlaid are detection significance contours of 3,4,5,6 $\sigma$ (derived from integration radii of again $0.1^{\circ}$). The ellipse indicates the outer boundary of the radio SNR candidate G323.7$-$1.0 detected in MGPS2 data \cite{bib:greenmgps22014}. Right: TeV surface brightness map of HESS\,J1912+101 derived analogously to the one for HESS\,J1534$-$571 (adopting in this case $\Gamma = 2.7$ from \cite{bib:1913hess2008}), and overlaid with detection significance contours of 3,4,5,6,7 $\sigma$. The insets on the lower left represent the resolution of the maps (i.e. the simulated instrumental point spread function of the respective data set, converted into a surface brightness with oversampling, and smoothed with the same function as the observational data).
}
	\label{Fig:SNRSkyMaps}
\end{figure}

\section{HESS\,J1534$-$571}

The grid search for shell-like sources as described in Sect.\,2 yields a mildly significant test significance (TS) difference between the two tested models (projected shell vs.\ 2D Gaussian) at Galactic coordinates $l\sim 323.6^{\circ}, b\sim-1.0^{\circ}$. The H.E.S.S.\ TeV source at that position has not been published previously, but the source is well above detection threshold following the detection procedure and threshold of the HGPS.\footnote{A test using a 2-dimensional Gaussian source hypothesis against the background hypothesis as used for the forthcoming HGPS catalog yields a test significance difference of $TS_{\mathrm{diff}} = 39$, well above HGPS detection threshold ($TS_{\mathrm{diff}} \geq 30$).} 

For the assessment of the morphology on individual source basis as also introduced in Sect.\,2, a dedicated analysis of the region around the new source has been performed. Data were analyzed with a similar analysis configuration as the HGPS primary analysis (Hillas-based event reconstruction and boosted decision trees for gamma-hadron separation \cite{bib:tmva2011}). All available observations within a distance of $3^{\circ}$ from the source position and acceptable observing conditions were used, resulting in a livetime-corrected\footnote{i.e. deadtime-corrected and corrected for acceptance loss at larger off-optical axis angles} exposure of $57.4\,\mathrm{h}$ at the source position. The TeV sky-map of HESS\,J1534$-$571 is shown in the left panel of Fig.\,\ref{Fig:SNRSkyMaps}. The signal-to-noise of the source excess (expressed by the Li\&Ma significance formula) is $9.3\,\sigma$, integrating excess within a radius of $0.38^{\circ}$ around the source centroid as defined through the center position and outer radius of a shell model fitted to this sky map. The name of the H.E.S.S. source (HESS\,J1534$-$571) has been chosen to match the centroid coordinates of the shell.

The statistical improvement when fitting a projected shell instead of a Gaussian null-hypothesis morphology to the sky map is quantified using the Akaike information criterion \cite{bib:AIC1974}, since this estimator permits the comparison of non-nested models such as the ones used. A null-hypothesis probability of $p=6.4\times10^{-3}$ for HESS\,J1534$-$571 is derived. Thus, from the TeV data alone the source has preliminarily been classified as a TeV shell candidate. However, in the course of the analysis, a set of new radio SNR candidates was published \cite{bib:greenmgps22014} using data from the Molonglo Galactic Plane Survey MGPS2. The newly detected radio SNR candidate G323.7$-$1.0 is in very good positional agreement with the H.E.S.S.\ source. In addition, the extension and the shell appearance match very well. From this spatial correlation and from the lack of any other known plausible counterparts, the source is classified as a TeV SNR. While still non-negligible, the chance probability that the source turns out to be e.g.\ a superbubble is considered sufficiently small.

The source has no counterpart in \emph{ROSAT} X-ray (survey) data. Due to the location of the source close to the Galactic plane, foreground absorption could have prevented a detection of the source in \emph{ROSAT} data, given its upper energy sensitivity limit at $\sim 2.4\,\mathrm{keV}$. The object has, however, also been the target of four \emph{Suzaku} observations (40\,ks each, PI A. Bamba) that partially covered the TeV source. No X-ray emission is found from the source region, excluding non-thermal X-ray emission at the level detected from the other known TeV SNR. Thus, the object is an excellent candidate for a TeV SNR whose TeV emission is dominated by proton-induced processes.

\section{HESS\,J1912+101}

HESS\,J1912+101 is a TeV source that has been discovered in 2008 \cite{bib:1913hess2008} during the ongoing H.E.S.S. survey. The source was published lacking firm identification, but with a plausible association with the energetic pulsar PSR J1913+1011 in a pulsar wind nebula scenario. However, no conclusive claim for this scenario was made, given the lack of a known PWN counterpart in other wavelengths that could support the offset PWN scenario, and, more importantly, lacking an energy-dependent TeV extension that would confirm the PWN identification; such successful spectro-morphological evidence is considered definitive proof of the TeV PWN natures of HESS\,J1825$-$137 \cite{bib:1825hess2006} and HESS\,J1303$-$631 \cite{bib:1303hess2012}.

Meanwhile, the H.E.S.S. exposure on the source has increased considerably, by a factor of $\sim 6$. This much larger data set changes the morphological picture of the TeV source significantly. The gridded shell search in the HGPS data as introduced in Sect.\,2 yields a high test significance difference at the position of HESS\,J1912+101. The assessment of the morphology on individual source basis was performed with a data set equivalent to the HGPS data set, using the same ana\-ly\-sis configuration as for HESS\,J1534$-$571 introduced above. The effecive livetime (defined as in Sect.\,3) amounts to $121.6\,\mathrm{h}$. A shell appearance of the source is now evident, as can be seen in the right panel of Fig.\,\ref{Fig:SNRSkyMaps}. The signal-to-noise of the source excess (expressed by the Li\&Ma significance formula) is $17.3\,\sigma$, integrating excess within a radius of $0.48^{\circ}$ around the source centroid as defined through the center position and outer radius of the shell model fitted to this sky map. The comparison of projected shell vs.\ Gaussian morphology yields a null-hypothethis probability of $p=1.7\times10^{-6}$, again using the Akaike information criterion \cite{bib:AIC1974}. Thus, the source is found to be a significant TeV shell and therefore a new SNR candidate.

In contrast to HESS\,J1534$-$571, there is no known radio counterpart that matches the TeV shell morphology. Given the more northern location of HESS\,J1912+101, it is not covered by the MGPS2 survey. Neither is it covered by the FIRST \cite{bib:firstcatalog1997} radio survey. A check of the NVSS and VLA GPS maps remains inconclusive. Given the low surface brightness of the other identified TeV SNRs, this finding is not considered a counterargument to the SNR scenario.

Concerning X-rays, the source has no \emph{ROSAT} counterpart. An excellent coincidental \emph{ASCA} coverage of the source yields only inconclusive results, due to strong straylight contamination stemming from GRS\,1915+10 \cite{bib:1913hess2008}. A \emph{Chandra} observation of the central region of the TeV source \cite{bib:changpwne2008} does not cover the newly detected shell; therefore, X-ray quietness of the shell (on current satellite sensitivity level) cannot be claimed.

\section{Brief discussion}

The TeV $\gamma$-ray detections of HESS\,J1731$-$347 \cite{bib:darkhess2008,bib:1731hess2011} and HESS\,J1534$-$571 (this work) with subsequent identification of radio SNR candidate counterparts demonstrate the capability of the current generation of TeV instruments to discover new SNRs. This gives further support to the hypothesis that HESS\,J1912+101 -- now a TeV SNR candidate without counterpart -- is indeed likely a TeV SNR; for the time being, the source is likely the first TeV SNR without known counterpart in other wavebands.

All {\it already established} TeV SNRs have non-thermally emitting X-ray SNR counterparts. An interpretation of the TeV $\gamma$-ray spectra in the framework of models yielding hadronically-dominated emission seems to work in many cases. However, scenarios in which the TeV emission stems from relativistic electrons emitting $\gamma$-rays via IC scattering present a very viable alternative interpretation, also supported by the data in the \emph{Fermi}-LAT $\gamma$-ray energy band (e.g.\ \cite{bib:17131006LAT2015}). Thus, while indirect evidence of proton acceleration exists from other observations, there is no conclusive proof that the TeV emission is tracing relativistic protons accelerated and/or confined in the shells.

Searching for (and finding) new TeV SNRs that have no non-thermal X-ray counterpart is thus an important method to detect proton-dominated TeV SNRs. The lack of a counterpart in \emph{ROSAT} data may not be conclusive given the likely strong foreground absorption towards the sources (cf.\ e.g.\ to HESS\,J1731$-$347 \cite{bib:1731hess2011}). Dedicated X-ray (re)-observations of the objects with current-generation or future X-ray satellites is therefore key towards establishing the nature of the relativistic particle population that gives rise to the TeV emission in these objects.

\section*{Acknowledgement}

The support of the Namibian authorities and of the University of Namibia in facilitating the construction and operation of H.E.S.S.\ is gratefully acknowledged, as is the support by the German Ministry for Education and Research (BMBF), the Max Planck Society, the German Research Foundation (DFG), the French Ministry for Research, the CNRS-IN2P3 and the Astroparticle Interdisciplinary Programme of the CNRS, the U.K.\ Science and Technology Facilities Council (STFC), the IPNP of the Charles University, the Czech Science Foundation, the Polish Ministry of Science and Higher Education, the South African Department of Science and Technology and National Research Foundation, and by the University of Namibia. We appreciate the excellent work of the technical support staff in Berlin, Durham, Hamburg, Heidelberg, Palaiseau, Paris, Saclay, and in Namibia in the construction and operation of the equipment.

The shell search performed on the entire HGPS data set as presented in section 2 employed Gammapy routines as presented in \cite{bib:GammapyICRC2015}.

\end{document}